# Improved Physics-based Raman Amplifier Model in C+L Networks through Input Parameter Refinement

Yihao Zhang, Xiaomin Liu, Qizhi Qiu, Yichen Liu, Lilin Yi, Weisheng Hu, and Qunbi Zhuge*

*State Key Laboratory of Advanced Optical Communication Systems and Networks, Department of Electronic Engineering, Shanghai Jiao Tong University, Shanghai 200240, China*
*Corresponding author e-mail address: qunbi.zhuge@sjtu.edu.cn*

**Abstract:** We propose an input parameter refinement scheme for the physics-based Raman amplifier model. Experiments over C+L band are conducted. Results show the scheme can lower the physical model's maximum estimation error by ~2.13 dB. © 2024 The Authors

## 1. Introduction

Accurate physical-layer modeling is crucial in building digital twins (DTs) for autonomous driving optical networks (ADONs) [1-2]. In practice, a major problem that deteriorates the precision of physical models is the uncertainty of input parameters. The causes of input parameter uncertainty are diverse, including measurement errors, discrepancies between the true and configured values, and aging degradation. To address this problem, efforts have been made to conduct input-parameter refinement (IR) [3-7]. These existing works mainly focus on refining the uncertainties of lumped loss and parameters of erbium-doped fiber amplifiers (EDFAs) in lumped-amplification scenarios.

Currently, the IR of physics-based Raman amplifier (RA) model has not received much attention in the existing literatures. The utilize of RA will become increasingly important with the development of ultra-wideband (UWB) systems [8-10]. Thus, it is crucial to study the accuracy of physics-based RA models in practical systems and focus on the IR of the RA model. A major challenge is the coupling of multiple parameters in Raman amplification. For instance, overestimating the received signal power can result from the overestimation of the pump power, the overestimation of Raman gain efficiency, and/or the underestimation of the attenuation. It is challenging to determine the extent of each over- or underestimation. Thus, gradient-based methods, which are adopted in lumped-amplification IR cases [3,5-7], are not well suited to this problem as the problem is non-convex. In [11], an evolutionary optimization strategy is utilized to characterize fiber parameters and Raman pump polarization coefficients, which produces satisfactory results and indicates the advantage of adopting a heuristic algorithm to this problem. However, the method is based on accurate measurement of Raman pump powers, which can be hard to achieve in real systems. Furthermore, the improvement of IR on the accuracy of physical Raman model based on ordinary differential equations (ODEs) has not been shown, and the investigation has only been carried out over the C band.

In this paper, an IR scheme for Raman-amplified links based on the particle swarm optimization (PSO) algorithm is proposed. Instead of only targeting to obtain true parameters under certain fixed working conditions, the proposed scheme can effectively learn the *mappings* from uncertain parameters to true parameters under various working conditions. The advantage is that the true parameters do not need to be re-obtained if the configurations (*e.g.*, the pump power or the signal power) change. The parameter uncertainties of signal power spectra, fiber attenuation, Raman gain efficiency, and Raman pump power are taken into consideration. Experiments of C+L band Raman amplification are conducted over a single span. Results show that the proposed IR scheme can lower the mean prediction error from ~1.22 dB to ~0.30 dB, and can lower the maximum prediction error from ~3.11 dB to ~0.98 dB.

## 2. The Proposed IR Scheme
*2.1 Uncertain Parameters and Assumptions*

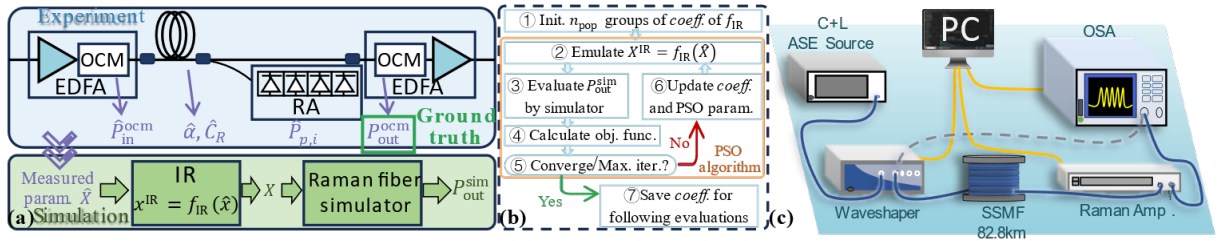

Fig. 1. **(a)** The parameters with uncertainties in a Raman-amplified span and an illustrative block diagram of IR. **(b)** The block diagram of proposed IR algorithm. **(c)** The experiment setup of this paper.

A single span with hybrid RA-EDFA amplification is depicted in Fig. 1(a). The RA is placed at the end of the fiber to provide backward distributed amplification. At the start and end of each span, the signal power spectrum is monitored. In commercial systems, this can be achieved by optical channel monitors (OCMs). We only focus on the model of RA in this work, so only the parameters between the two OCMs shown in the figure are considered. The monitored signal power spectrum at the end of the span, denoted as $P_{\text{out}}^{\text{ocm}}$, is chosen as the *ground truth*. In other words, we assess the accuracy of the physics-based ODE model by comparing $P_{\text{out}}^{\text{ocm}}$ with $P_{\text{out}}^{\text{sim}}$, which refers to the received power spectrum calculated by the Raman ODE solver in simulations.

According to the Raman ODE [8, Eq.(1-2)], parameters that affect the calculation of $P_{\text{out}}^{\text{sim}}$ include the number of signals and pumps, the frequencies of all signals and pumps, the length of the fiber, the pump powers ($P_{p,i}$), the signal power spectrum launched into the fiber ($P_{\text{in}}^{\text{ocm}}$), the attenuation profile of the fiber ($\alpha$), and the Raman gain efficiency profile of the fiber ($C_R$). In this work, it is assumed that all the frequencies and the fiber length are accurate, whereas other parameters contain uncertainties. The aim of the proposed scheme is to find the mappings $f_{\text{IR}}$ from the measured/acquired parameters $\hat{x}$ to the true parameters $x^{\text{IR}}$:

$$x^{\text{IR}} = f_{\text{IR}}(\hat{x}), \quad (1)$$

where $x^{\text{IR}} \in X = \{x|x = P_{p,i}^{\text{IR}}, P_{\text{in}}^{\text{IR}}, P_{\text{out}}^{\text{sim}}, \alpha^{\text{IR}}, C_R^{\text{IR}}\}$ and $\hat{x} \in \hat{X} = \{x|x = \hat{P}_{p,i}, \hat{P}_{\text{in}}^{\text{ocm}}, \hat{P}_{\text{out}}^{\text{sim}}, \hat{\alpha}, \hat{C}_R\}$.

Different types of functions are chosen for $f_{\text{IR}}$ of different parameters, which are detailed as follows. First, for pump powers $P_{p,i}$, it is assumed that $P_{p,i}^{\text{IR}}(\text{W}) = m \cdot \hat{P}_{p,i}(\text{W}) + n$. We have found that such a simple linear function can effectively fit the inaccuracies in the pump power. Second, for signal power spectra $P_{\{\text{in,out}\}}$, it is assumed that $P_{\text{in}}^{\text{IR}}(\text{dBm}) = \hat{P}_{\text{in}}^{\text{ocm}}(\text{dBm}) - \delta_1$ and $P_{\text{out}}^{\text{sim}}(\text{dBm}) = \hat{P}_{\text{out}}^{\text{sim}}(\text{dBm}) - \delta_2$. The coefficients $\delta_1$ and $\delta_2$ represent the unknown lumped loss due to connections and the measurement error of OCMs. Last, for fiber attenuation $\alpha$ and Raman gain efficiency $C_R$, we assume that their curve shapes obtained from the datasheet are correct but with some deviations, which are defined as $\alpha^{\text{IR}} = a \cdot \hat{\alpha}$ and $C_R^{\text{IR}} = c \cdot \hat{C}_R$. As a result, six $f_{\text{IR}}$ coefficients, denoted as $\Gamma = (m, n, \delta_1, \delta_2, a, c)$, need to be figured out through the IR algorithm. $N_{\text{train}}$ groups of parameters and corresponding ground truth $P_{\text{out}}^{\text{ocm}}$ are collected in advance for the searching of these $f_{\text{IR}}$ coefficients.

*2.2 PSO-based IR Algorithm Description*

The PSO algorithm is utilized for searching the optimal $\Gamma$ owing to its simplicity and fast convergence speed. The block diagram is shown in Fig. 1(b). First, $n_{\text{pop}}$ particles are generated. $n_{\text{pop}}$ groups of randomly generated $\Gamma$ are allocated to the particles as their initial positions. Their velocities are also randomized. Second, for each particle, the estimated true value set $X^{\text{IR}}$ is calculated. Then $P_{\text{out}}^{\text{sim}}$ is evaluated by the Raman ODE solver. Next, the objective function is calculated, defined as the maximum error between $P_{\text{out}}^{\text{sim}}$ and $P_{\text{out}}^{\text{ocm}}$ among all $N_{\text{train}}$ data and $N_{\text{ch}}$ channels:

$$f_{\text{obj}}(\Gamma) = \max_{j=1,\cdots,N_{\text{train}}} \left\{ \max_{k=1,\cdots,N_{\text{ch}}} \left| P_{\text{out},j,k}^{\text{sim}} - P_{\text{out},j,k}^{\text{ocm}} \right| \right\}, \quad (2)$$

whose value is the fitness of each particle. The particle's position is considered better as the fitness value decreases. Afterwards, we check if the termination condition is met, such as the minimal fitness remaining unchanged for $n_{\text{patience}}$ iterations, or reaching the maximum number of iterations. If the condition is met, stop the algorithm. Otherwise, the positions (i.e., $\Gamma$) and velocities of all particles are updated according to the requirements of the PSO algorithm [12]. Then a next round of iteration goes on.

## 3. Experiment Setup

Experiments are conducted as depicted in Fig. 1(a), and the setup is shown in Fig. 1(c). A C+L amplified spontaneous emission (ASE) source providing an 8.4-THz-wide spectrum ranging from 187.6 THz to 196.0 THz is utilized to emulate 42 signal channels with a grid of 200 GHz. A programmable optical filter is connected afterwards to adjust the shape of signal spectrum. Then the signals are transmitted over an 82.8-km long G.652D standard single-mode fiber. An RA is employed at the end of the fiber to provide counter-directional Raman amplification, with four Raman pumps whose wavelengths are 1428 nm, 1454 nm, 1490 nm, and 1509 nm. The pump powers can be adjusted by controlling the digital-to-analog convertor (DAC) setting values. An optical spectrum analyzer (OSA) is connected to the output of the programmable optical filter or the output of the RA to emulate the OCMs.

First, the fiber attenuation profile is measured by filtering out the signal within each 200-GHz channel and monitoring the power attenuation by the OSA. This approach avoids the effect of inter-channel stimulated Raman scattering (SRS) on the measurements of attenuation profile. Then, the pump powers are measured by an optical power meter under different DAC values. A linear function is fitted to characterize the relationship between the set DAC values and pump powers (in watt), which is $P_{p,i} = 0.00015\text{DAC}_i - 0.01916$. The initial $\hat{C}_R$ profile is obtained from the device datasheet.

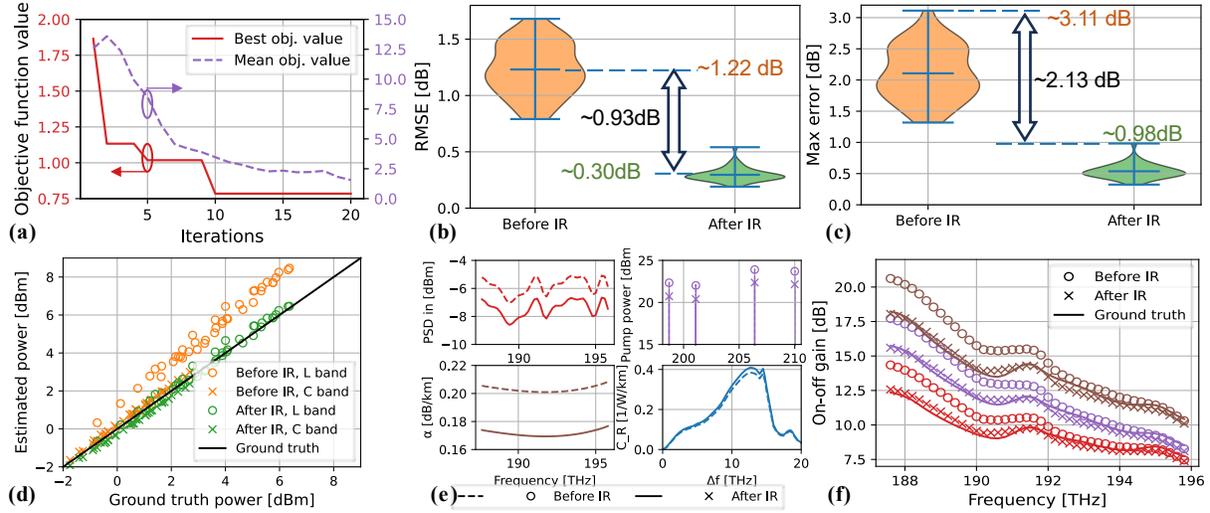

Fig. 2. **(a)** The convergence curve of the PSO algorithm. **(b)** The RMSE of received signal profile estimation before and after IR. **(c)** The maximum error of received signal profile estimation before and after IR. **(d)** C-band and L-band power estimation before and after IR. **(e)** The refined parameters. **(f)** Three on-off gain profiles estimated by simulation before and after IR, and the corresponding ground truth profiles.

Next, $N_{train} = 10$ groups of DAC values (from 800 to 2000) are randomly generated and configured for the 4 pumps to emulate 10 different working conditions. Under each condition, the launch and received signal spectra are measured. Using the same method, another $N_{test} = 40$ groups of data are collected for subsequent validation on the physics-based model's accuracy after IR. During the PSO algorithm, we let $n_{pop} = 32$ and $n_{patience} = 10$.

## 4. Results

First, the variance of the objective function value during optimization is shown in Fig. 2(a). The mean objective function value of all particles continues to descend, which illustrates that the algorithm converges well. After 10 iterations, the optimal coefficient set appears to be $\Gamma_{opt} = (1.574, 0.182, -0.032, 1.065, 0.728, -0.006)$. Then, using $\Gamma_{opt}$ to conduct IR, the physics-based Raman ODE model is validated on the collected 40 test data. The results are shown in Fig. 2(b) and 2(c). In Fig. 2(b), the distributions of root mean square errors (RMSEs) between $P_{out}^{sim}$ and $P_{out}^{ocm}$ before and after IR are plotted. It is shown that the mean of RMSE over the test-set can be reduced by ~0.93 dB, from ~1.22 dB to ~0.30 dB. Similarly, in Fig. 2(c), the maximum error between $P_{out}^{sim}$ and $P_{out}^{ocm}$ before and after IR are compared. Results show that the maximum estimation error can be reduced by ~2.13 dB, from ~3.11 dB to ~0.98 dB. Furthermore, in Fig. 2(d), the estimated power versus the ground truth power of C band and L band is shown. It can be observed that the estimated power of L band is evidently higher than the ground truth before IR. This discrepancy can be significantly reduced by IR.

To clearly present the accuracy improvement provided by IR, an example of parameters before and after IR is plotted in Fig. 2(e), and 3 examples of estimated on-off gain profiles before and after IR are plotted in Fig. 2(f). It can be found that before IR, the launch signal power, the pump power, and the attenuation is overestimated, whereas the Raman gain efficiency is slightly underestimated. This results in the overestimation on the on-off gain, especially for L band, as shown in Fig. 2(f). Through IR, the estimated on-off gain profiles can closely match the ground truth. To sum up, the proposed IR scheme can effectively reduce the estimation error of the physics-based ODE model of RA.

## 5. Conclusions

A PSO-based input parameter refinement scheme for the physics-based RA ODE model is proposed. The mappings from the uncertain power spectrum, Raman pump power, fiber attenuation, and Raman gain efficiency to their true values can be fitted by the proposed IR scheme. Through experiments in a C+L system, results show that the proposed IR scheme can reduce the mean signal spectrum estimation error by ~0.93 dB and the maximum error by ~2.13 dB.

*Acknowl.:* supported by the Shanghai Pilot Program for Basic Research - Shanghai Jiao Tong University (21TQ1400213) and NSFC (62175145)